\documentclass{sigchi}

\usepackage{balance}  
\usepackage{graphics} 
\usepackage{url}      
\usepackage{latexsym}
\usepackage{amsfonts, amsmath}
\usepackage{multirow}
\usepackage{color}
\usepackage{enumerate}
\usepackage{comment}
\usepackage{bbm}
\usepackage{xspace}
\usepackage{booktabs}
\usepackage{enumitem}

\usepackage{times}    

\makeatletter
\def\url@leostyle{%
  \@ifundefined{selectfont}{\def\UrlFont{\sf}}{\def\UrlFont{\small\bf\ttfamily}}}
\makeatother
\def\pprw{8.5in}
\def\pprh{11in}

\newcommand{\ourtitle}{SceneSeer: 3D Scene Design with Natural Language}

\usepackage[pdfencoding=auto]{hyperref}
\hypersetup{
pdftitle={\ourtitle},
pdfauthor={Angel X. Chang, Mihail Eric, Manolis Savva, Christopher D. Manning},
pdfkeywords={},
bookmarksnumbered,
pdfstartview={FitH},
colorlinks,
citecolor=black,
filecolor=black,
linkcolor=black,
urlcolor=black,
breaklinks=true,
}

\newcommand{\stt}[1]{\small{\texttt{#1}}}

\newcommand{\cbasic}{\texttt{basic}\xspace}
\newcommand{\csup}{\texttt{+sup}\xspace}
\newcommand{\csupspat}{\texttt{+sup+spat}\xspace}
\newcommand{\csupprior}{\texttt{+sup+prior}\xspace}
\newcommand{\cfull}{\texttt{full}\xspace}
\newcommand{\cinfer}{\texttt{full+infer}\xspace}
\newcommand{\cmanual}{\texttt{human}\xspace}
\newcommand{\SceneSeer}{\textsc{SceneSeer}\xspace}

\DeclareMathOperator*{\argmax}{arg\,max}

\title{\ourtitle}

\numberofauthors{1}
\author{
  \alignauthor Angel X. Chang, Mihail Eric, Manolis Savva, \textnormal{\large and} Christopher D. Manning\\
  \affaddr{Computer Science Department, Stanford University}\\
  \email{$\{\textnormal{\large angelx, meric, msavva, manning\}@cs.stanford.edu}$}\\
}
\toappear{}
\teaser{
  \includegraphics[width=\linewidth]{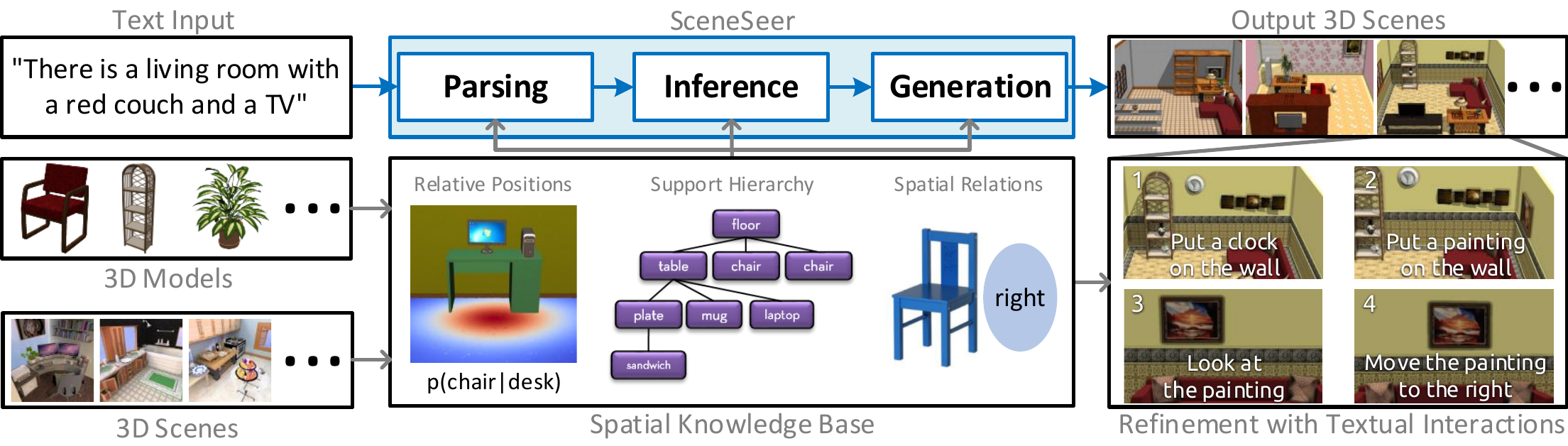}
  \caption{Illustration of the \SceneSeer architecture.  Input text is parsed into a scene representation and expanded through inference using knowledge learned from a corpus of 3D scenes and 3D models.  The representation is then used to generate a 3D scene that can be rendered and manipulated iteratively through textual commands.}
  \label{fig:teaser}
}

\begin{document}

\maketitle

\begin{abstract}
Designing 3D scenes is currently a creative task that requires significant expertise and effort in using complex 3D design interfaces.  This effortful design process starts in stark contrast to the easiness with which people can use language to describe real and imaginary environments.  We present \SceneSeer: an interactive text to 3D scene generation system that allows a user to design 3D scenes using natural language.  A user provides input text from which we extract explicit constraints on the objects that should appear in the scene.  Given these explicit constraints, the system then uses a spatial knowledge base learned from an existing database of 3D scenes and 3D object models to infer an arrangement of the objects forming a natural scene matching the input description.  Using textual commands the user can then iteratively refine the created scene by adding, removing, replacing, and manipulating objects.  We evaluate the quality of 3D scenes generated by \SceneSeer in a perceptual evaluation experiment where we compare against manually designed scenes and simpler baselines for 3D scene generation.  We demonstrate how the generated scenes can be iteratively refined through simple natural language commands.
\end{abstract}

\section{Introduction}

Designing 3D scenes is a challenging creative task.  Expert users expend considerable effort in learning how to use complex 3D scene design tools. Still, immense manual effort is required, leading to high costs for producing 3D content in video games, films, interior design, and architectural visualization.  Despite the conceptual simplicity of generating pictures from descriptions, systems for text-to-scene generation have only achieved limited success.  How might we allow people to create 3D scenes using simple natural language?

Current 3D design tools provide a great amount of control over the construction and precise positioning of geometry within 3D scenes.  However, most of these tools do not allow for intuitively assembling a scene from existing objects which is critical for non-professional users.  As an analogue, in real life few people are carpenters, but most of us have bought and arranged furniture.  For the purposes of defining how to compose and arrange objects into scenes, natural language is an obvious interface.  It is much easier to say ``Put a blue bowl on the dining table'' rather than retrieving, inserting and orienting a 3D model of a bowl.  Text to 3D scene interfaces can empower a broader demographic to create 3D scenes for games, interior design, and virtual storyboarding.

Text to 3D scene systems face several technical challenges.  Firstly, natural language is typically terse and incomplete.  People rarely mention many facts about the world since these facts can often be safely assumed.  Most desks are upright and on the floor but few people would mention this explicitly.  This implicit spatial knowledge is critical for scene generation but hard to extract.  Secondly, people reason about the world at a much higher level than typical representations of 3D scenes (using the descriptive phrase \emph{table against wall} vs a 3D transformation matrix).  The semantics of objects and their approximate arrangement are typically more important than the precise and abstract properties of geometry.  Most 3D scene design tools grew out of the traditions of Computer Aided Design and architecture where precision of control and specification is much more important than for casual users.  Traditional interfaces allow for comprehensive control but are typically not concerned with high level semantics.

\SceneSeer allows users to generate and manipulate 3D scenes at the level of everyday semantics through simple natural language. It leverages spatial knowledge priors learned from existing 3D scene data to infer implicit, unmentioned constraints and resolve view-dependent spatial relations in a natural way.  For instance, given the sentence ``there is a dining table with a cake'', we can infer that the cake is most likely on a plate and that the plate is most likely on the table.  This elevation of 3D scene design to the level of everyday semantics is critical for enabling intuitive design interfaces, rapid prototyping, and coarse-to-fine refinements.

In this paper, we present a framework for the text to 3D scene task and use it to motivate the design of the \SceneSeer system.  We demonstrate that \SceneSeer can be used to generate 3D scenes from terse, natural language descriptions.  We empirically evaluate the quality of the generated scenes with a human judgment experiment and find that \SceneSeer can generate high quality scenes matching the input text.  We show how textual commands can be used interactively in \SceneSeer to manipulate generated 3D scenes.

\section{Background}

\subsection{3D Design Interfaces}

Current research typically focuses on low-level interfaces for 3D modeling.  There is much work on input methodologies and schemes for easing navigation in 3D~\cite{khan2008viewcube,mccrae2009multiscale} and manipulation of primitives such as curves~\cite{owen2005gets}.  However, there is little work on 3D scene manipulation at a semantic level.  Recent work has explored the avenue of parsing natural language for design problem definitions in the context of mechanical computer-aided design software~\cite{cheong2014natural}.  Our motivation is similar, but we focus on the more open-ended setting of natural language 3D scene generation and manipulation.


Another line of work focuses on defining higher-level semantic APIs for scripting 3D animations and storyboards~\cite{conway2000alice,kelleher2006lessons}.  This prior work finds that people overwhelmingly prefer higher-level manipulation and specification in terms of semantic concepts such as ``in front'', ``to the left'' and ``on top of'' rather than low level manipulation.  The focus of this work was on defining simple APIs for designing animated stories, whereas we focus on demonstrating a natural language interface for static 3D scene design.  We are closer to early seminal work in placing simple geometric objects in 3D through textual commands~\cite{clay1996put}.

More recently, Chaudhuri et al.~\shortcite{chaudhuri2013attribit} have demonstrated a novel 3D model design interface where users can control the desired values of high-level semantic attributes (e.g., ``aerodynamic'', ``scary'') to interactively combine object parts.  We similarly allow users high-level semantic control but our focus is on 3D scenes where manipulating the layout of objects comes with a different set of challenges compared to object assembly.

\subsection{Automatic Scene Layout}

Recent research in computer graphics has focused on learning how to automatically lay out 3D scenes given training data.  Prior work on scene layout has focused largely on room interiors and determining good furniture layouts by optimizing energy functions that capture the quality of a proposed layout.  These energy functions are encoded from interior design guidelines~\cite{merrell2011interactive} or learned from input scene data~\cite{fisher2012example}.  Knowledge of object co-occurrences and spatial relations is represented by simple models such as mixtures of Gaussians on pairwise object positions and orientations. Methods to learn scene structure have been demonstrated using various data sources including simulation of human agents in 3D scenes~\cite{jiang2012learning,jiang2013infinite}, and analysis of supporting contact points in scanned environments~\cite{rosman2011learning}.  Though \SceneSeer is also a system for generating 3D scenes, it focuses on providing an interactive natural language interface for this task.

\subsection{Text to Scene Systems}

Early work in textual interfaces for 3D scene manipulation has addressed simplified scenarios with micro-worlds consisting of simple geometric shapes.  The SHRDLU~\cite{winograd1972understanding} and PUT~\cite{clay1996put} systems were pioneers in parsing natural language instructions, but generalization of their approach to more realistic scenarios similar to the real world is challenging.  More recent work on the WordsEye system~\cite{coyne2001wordseye,coyne2012annotation} and other similar approaches~\cite{seversky2006real} has demonstrated more general text-driven 3D scene generation that can handle complex scenes.  The authors compellingly show the benefit of text to scene generation but note that their systems are restricted due to a lack of implicit spatial knowledge.  As a result, unnatural language, such as ``the stool is 1 feet to the south of the table',' must be used to fully specify the scene.

Most recently, Chang et al.~\cite{chang2014spatial} have demonstrated how linguistic and non-linguistic spatial knowledge can be learned directly from existing scene data and leveraged when parsing natural language.  We build upon this prior work and focus on interactive scene design where the user does not have to specify all details of the scene at once, in contrast to prior text to scene systems such as WordsEye.  With \SceneSeer the user can dynamically manipulate a 3D scene with simple textual commands resembling dialogue systems.

\section{Approach Overview}

How might we create an interactive text to 3D scene system?  The user should be able to describe and manipulate a scene with concise natural language.  To make this possible, we need to parse the input text to a set of explicitly provided constraints on objects and their arrangement.  We also need to expand these constraints with implicit ``common sense'' facts (e.g., most of the time plates go on tables, not on the floor).  Once we've inferred implicit constraints, we can generate a candidate scene that can be rendered and viewed by the user for further interactive refinement with textual commands.  Based on the systems presented in \cite{chang2014interactive,chang2014spatial}, we frame the interactive text-to-scene problem in a probabilistic formulation that covers both scene generation and interactive scene operations. We show how these two previous systems can be viewed as specific instantiations of our framework.

\SceneSeer relies on a spatial knowledge base that is learned from a corpus of 3D scenes and 3D models (see Figure~\ref{fig:teaser}).  The learning procedure and components of this spatial knowledge base are based on the prior work of Chang et al.~\cite{chang2014spatial} and we use the same corpus of 3D scenes and component 3D models, consisting of about 12,490 mostly indoor objects with associated textual categories and tags.  Our approach for extracting object relative positions, support hierarchies, support surfaces, and spatial relations is also based on that of Chang et al., and is described in more detail in the appendix.

We define our problem as the task of taking text describing a scene as input, and generating a plausible 3D scene described by that text as output.  More concretely, based on the input text, we select objects from a dataset of 3D models and arrange them to generate output scenes.  See Figure~\ref{fig:teaser} for an illustration of the system architecture. We break the system down into several subtasks:

\textbf{Semantic Parsing:} Parse the input textual description of a concrete scene into a scene template consisting of constraints on the objects present and spatial relations between them.

\textbf{Scene Inference:} Automatically expand the scene template to account for implicit constraints not specified in the text.

\textbf{Scene Generation:} Using the above scene template and prior knowledge on the spatial arrangement of objects, sample the template, select a set of objects to be instantiated, and optimize their arrangement into a scene.

\textbf{Scene Interaction:} Allow user to interactively adjust the scene through direct manipulation and textual commands.

This breakdown is useful for tackling the hard problem of text-to-scene and allows us to address subproblems independently or jointly.  We choose this breakdown as it provides a separation between subproblems that fall within the graphics, NLP, and interaction design communities.

\subsection{Representation}

There is a big gap between the representations typically used for 3D object and scene rendering, and the high-level semantics that people assign to scenes.  Here we define a scene template representation following Chang et al.~\cite{chang2014spatial}'s approach to make explicit the information necessary for connecting higher-level semantics to the lower-level geometric representations.  Since natural language usually expresses high-level semantics, we can view the text-to-scene task as a problem of first extracting a higher-level scene template representation and then mapping into a concrete geometric representation.

\paragraph{Scene Template}
A scene template $t=(\mathcal{O}, \mathcal{C}, C_s)$ consists of a set of object descriptions $\mathcal{O} = \{o_1, \ldots, o_n\}$ and constraints $\mathcal{C} = \{c_1, \ldots, c_k\}$ on the relationships between the objects. A scene template also has a scene type $C_s$.  This provides a high-level yet unambiguous representation of scene structure.

Each object $o_i$, has properties associated with it such as category label, basic attributes such as color and material, and number of occurrences in the scene.  For constraints, we focus on spatial relations between objects, expressed as predicates of the form $\textit{supported\_by}(o_i,o_j)$ or $\textit{left}(o_i,o_j)$ where $o_i$ and $o_j$ are recognized objects.

\paragraph{Geometric Scene}
We refer to the concrete geometric representation of a scene as a ``geometric scene''. A geometric scene $s$ consists of a set of 3D model instances $\{i_1,\ldots,i_n\}$ where each model instance $i_j = (m_j, T_j)$ is represented by model $m_j$  in the model database and the transformation matrix $T_j$.  The model represents the physical appearance (geometry and texture) of the object while the transformation matrix encodes the position, orientation, and scaling of the object.  Working directly with such low-level representations of scenes is unnatural for people, which is a factor in the difficulty of learning current 3D scene design interfaces.  We generate a geometric scene from a scene template by selecting appropriate models from a 3D model database and determining transformations that optimize their layout to satisfy spatial constraints.

\subsection{Model Formulation}
We take a probabilistic view and model the task as that of estimating the distribution of possible scenes $P(s|u)$ given the input utterance $u$.  This allows us to incorporate prior knowledge to perform inference, and to leverage recent advances in machine learning for learning from data.  The distribution $P(s|u)$ can be sampled from to generate plausible scenes to present to the user.

We can further decompose $P(s|u)$ into:
$$
P(s|u) = P(t|u) P(t'|t,u) P(s|t',t,u)
$$
where $P(t|u)$ is the probability of a scene template $t$ given a utterance $u$, and $t'$ is a completed scene template.  In our pipelined system, we will assume that $s$ is independent of $t,u$ and that $t'$ is independent of $u$ to get
$$
P(s|u) = P(t|u) P(t'|t) P(s|t')
$$
A more sophisticated system would jointly estimate $P(s|u)$.

In our framework, \textbf{Semantic Parsing} corresponds to estimating $P(t|u)$, \textbf{Scene Inference} to estimating $P(t'|t)$, and \textbf{Scene Generation} to estimating $P(s|t')$.  The deterministic parsing model of Chang et al.~\shortcite{chang2014spatial} represents $P(t|u)$ as a delta probability at the selected $t$.

We decompose scene generation into two parts: \textbf{Object Selection} and \textbf{Scene Layout}.  Object selection consists of identifying a likely set of models $\{m_j\}$ from our model database given a complete scene template $t$.  Scene layout is the problem of arranging the objects and finding appropriate transforms $\{T_j\}$ for the models $\{m_j\}$ and scene template $t$.
$$
P(s|t) = P(\{m_j\}|t) P(\{m_j, T_j\}|\{m_j\},t)
$$
Again, a more sophisticated system may choose to estimate $P(s|t)$ jointly. Note that in the equation above, $\{m_j\}$ denotes the set of all models and $\{m_j, T_j\}$ denotes the set of all model, transformation matrix tuples $(m_j, T_j)$. 

To further simplify the problem, we assume that we can select the model for each object independently.  Thus, we have
$$
P(\{m_j\}|t) = \prod_{j} P(m_j|o_j)
$$
This assumption clearly does not hold for many scenes and scene templates (e.g., matching furniture sets).

For scene layout, we estimate $P(\{m_j, T_j\}|\{m_j\},t)$ using a layout score $\mathcal{L}$ described in more detail in the Appendix.

To support \textbf{Scene Interactions}, we follow a similar model, except now we are given a starting scene $s$ and an utterance $u$, and we want to determine the desired scene $s'$.  We can model this as estimating $P(s'|s,u)$.  For simplicity, we will assume that the utterance consist of one scene operation $O$ that takes $s$ and transforms it to $s'$.  Thus we can break down $P(s'|s,u) = P(s'|O,s,u) P(O|s,u)$.  Here, given a specific starting scene $s$ and the utterance $u$ expressing a desired operation, we want to find the operation $O$ that can be applied on $s$ to yield $s'$.  This problem can be decomposed into parsing, inference and generation steps that we use for regular scene generation as well.  The difference is that we now estimate an operation $O$.


\section{Semantic Parsing}

During semantic parsing we take the input text and create a scene template that identifies the objects and the relations between them.  We follow the approach of Chang et al.~\cite{chang2014spatial} to process the input text using the Stanford CoreNLP pipeline\footnote{\scriptsize\url{http://nlp.stanford.edu/software/corenlp.shtml}}, identify visualizable objects, physical properties of the objects, and extract dependency patterns for spatial relations between the objects.  The parsed text is then deterministically mapped to the scene template or scene operations.  An interesting avenue for future research is to automatically learn how to map text using more advanced semantic parsing methods.

\section{Scene Inference}

After we obtain a scene template with the explicitly stated constraints, we expand it to include inferred objects and implicit constraints.  As an example, for each object in the scene, we use the support hierarchy prior $P_\textit{support}$ to find the most likely supporting parent object category.  If there is no such parent object already in the scene, we add it to the scene.  For instance, given the objects ``computer'' and ``room'', we infer that there should also be a ``desk'' supporting the ``computer''.

In addition to inferring missing objects, we infer the static support hierarchy for the objects in the scene.  For each object, we first identify support constraints that must exist based on the scene template.  For any objects that do not have a supporting parent, we sample  $P_\textit{support}(C_p|C_c)$ restricted to the set of available object categories.  In addition, we can infer missing objects based on the location (for instance, a bedroom typically contains a bed and a dresser).  Future work would involve enhancing this component with improved inference by identifying what objects must exist for each location, instead of sampling a random selection of objects.  We can also consider additional contextual details such as the distinction between a messy desk with a variety of items versus a tidy desk with few items.

\begin{figure}
  \includegraphics[width=\columnwidth]{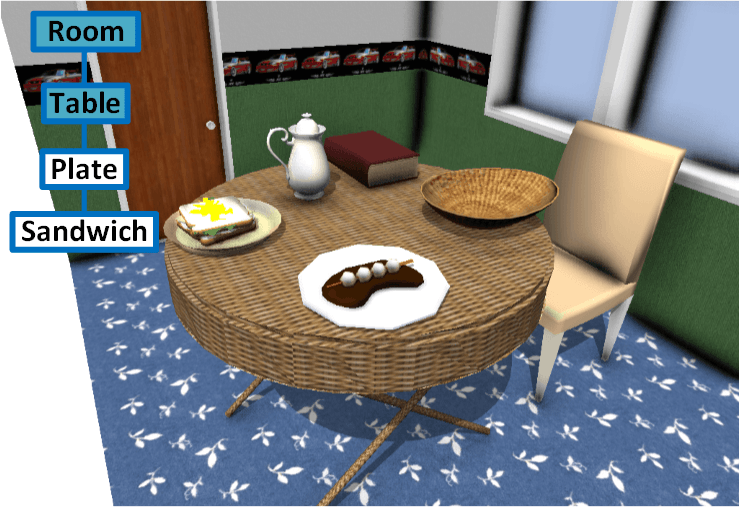}
  \caption{Generated scene for ``There is a sandwich on a plate.''  Note that the room, table and other objects are not explicitly stated, but are inferred by the system.}
  \label{fig:generatedSandwich}
\end{figure}

\begin{figure*}
  \includegraphics[width=\linewidth]{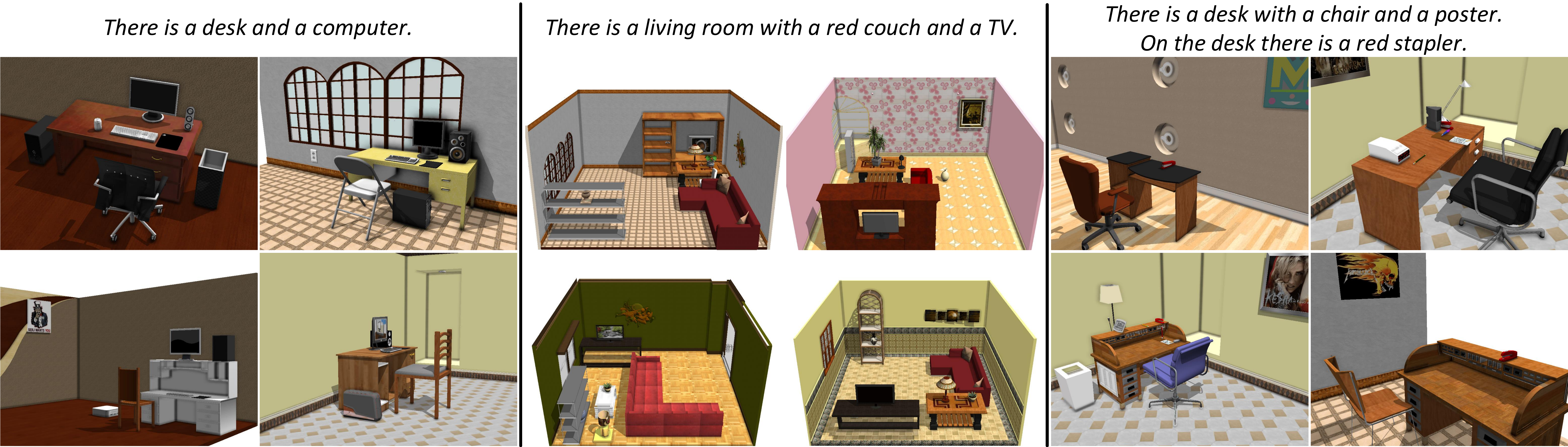}
  \caption{Several example generated scenes for three different input descriptions.}
  \label{fig:generatedScenes}
\end{figure*}

\begin{table*}
  \centering
  \small
  \begin{tabular}{lllll}
    Verb          & Operation &                  Example text              &                  Example parse           &  Influenced parameters \\ \toprule
    select         &  Select   & select the chair on the right of the table & $Select$(\{lamp\},\{right(lamp,table)\}) &  $\textit{Sel}$ \\
    look,look at      &  LookAt   &              look at the lamp              &            $LookAt$(\{lamp\})            &  $\textit{Cam}$ \\
    add,insert,place,put &  Insert   &          add a lamp to the table           &  $Insert$(\{lamp\},\{on(lamp,table)\})   &  $+\textit{object}$ \\
    delete,remove      &  Remove   &              remove the lamp               &            $Remove$(\{lamp\})            &  $-\textit{object}$ \\
    replace         &  Replace  &        replace the lamp with a vase        &       $Replace$(\{lamp\},\{vase\})       &  $+\textit{object},-\textit{object}$ \\
    move, place, put    &  Move     &         move the chair to the left         &    $Move$(\{chair\},\{left(chair)\})     &  $\textit{pos},\textit{parent}_\textit{sup},\textit{surf}_\textit{sup}$ \\
    enlarge, shrink     &  Scale    &         enlarge the bowl                   &    $Scale$(\{bowl\})(1.5)                &  $\sigma$ \\
  \end{tabular}
  \caption{Scene operations defined for our system.  The natural language verbs are parsed to specific operations and subsets of the scene parameters are modified accordingly (last column).}
  \label{tab:sceneOperations}
\end{table*}

\section{Scene Generation}
\label{sec:sceneGeneration}
Once we have a complete scene template we need to select 3D models to represent the objects trying to match any described attributes, and to arrange the models in the scene based on constraints.  During this step we aim to find the most likely scene given the scene template and prior spatial knowledge.

\textbf{Object Selection}
We use the keywords associated with each object to query the model database.  We select randomly from the top 10 results for variety and to allow the user to regenerate the scene with different models.  This step can be enhanced to take into account correlations between objects (e.g., a lamp on a table should not be a floor lamp model).

\textbf{Object Layout}
Given the selected models, the source scene template, and priors on spatial relations, we find an arrangement of the objects within the scene that maximizes the probability of the layout under the given scene template.

In order to compute how well the layout satisfies the constraints given by the scene template, we parameterize the layout of each object using semantically meaningful attributes: support parent $\textit{parent}_{sup}$, support surface $\textit{surf}_\textit{sup}$, attachment surface $\textit{surf}_\textit{att}$, position on support surface $\textit{pos}_\textit{sup}$, orientation $\theta$, size $\sigma$. These semantic attributes allow us to do higher level reasoning and interactions with the object.


We use a sampling approach to determine the position, orientation, and size of objects within the scene.  We first traverse the support hierarchy in depth-first order, positioning the largest available child node and recursing.  Child nodes are positioned by first selecting a supporting surface $\textit{surf}_\textit{sup}$ on a candidate parent object through sampling of $P_{\textit{surf}_\textit{sup}}$.  Using $P_{\textit{surf}_\textit{att}}$, we find the most appropriate child attachment surface $\textit{surf}_\textit{att}$ and orient the child node accordingly.  We then sample possible placements $(\textit{pos}_\textit{sup}, \theta, \sigma)$ on $\textit{surf}_\textit{sup}$, ensuring that the node is not overhanging and there are no collisions with other objects.  Each sampled placement is scored with $\mathcal{L}$. The resulting scene is rendered and presented to the user.


Figure~\ref{fig:generatedSandwich} shows a generated scene along with the support hierarchy and input text.  Even though the room, table, and other objects were not explicitly mentioned in the input, we infer that the plate is likely to be supported by a table and that there likely to be other objects on the table.  Without this inference, the user would need to be much more verbose with text such as ``There is a room with a table, a plate, and a sandwich.  The sandwich is on the plate, and the plate is on the table.''  

Figure~\ref{fig:generatedScenes} shows several examples illustrating how we can generate many 3D scenes for the user to select from.  Our system can infer many unstated but implicitly present objects (e.g., monitors are placed on the desks for functional computer setups).  We can generate high quality scenes for localized arrangements such as the computer desks, for which there are many examples in the training 3D scene corpus.  However, more complex scenes such as the living rooms exhibit some unnatural object positioning (e.g., TV does not face the couch).  For such cases, it is important to allow users to iteratively refine the generated scene.

\section{Scene Interactions}
\label{sec:sceneOperations}

Once a scene is generated, the user can view the scene and manipulate it using both textual commands and mouse interaction.  The system supports traditional 3D scene interaction mechanisms such as navigating the viewpoint with mouse and keyboard, and selection and movement of object models by clicking.  In addition, a user can give simple textual commands to select and modify objects, or to refine the scene.

To allow the user to interact with a generated scene via text, we define a set of high-level semantic operations for scene manipulation.  These are higher level operations than typically found in traditional scene editing software.  Again, the text is treated as a set of constraints that we want the revised scene to satisfy, while trying to keep the revised scene as similar to the original scene as possible.  More formally, given the original scene $S$ and a scene operation $O$, we want to find the scene $S'$ which is most similar to $S$, while still satisfying the constraints imposed by $O$.

To track the other elements of the scene, we maintain a scene state $Z = (S,\textit{Sel},\textit{Cam})$ that consists of the scene $S$, the set of selected objects $\textit{Sel}$, and the camera position $\textit{Cam}$.  Each operation can be defined as a function $O: Z \rightarrow Z'$.  We support several basic operations: $Select$ changes the set of selected objects, $Insert$ adds objects into the scene, $Delete$ removes objects from the scene, $Replace$ replaces objects in the scene with new objects, operations such as $Move$ and $Scale$ modify constraints on existing objects without changing the set of objects in the scene.  $LookAt$ repositions the camera to focus on the selected objects.  Table \ref{tab:sceneOperations} summarizes the operations supported by our system.  These basic operations demonstrate some simple scene manipulations through natural language.  This set of operations can be extended, for example, to cover manipulation of parts of objects (``make the seat of the chair red'').

To interpret a textual scene interaction command, the system first parses the input text $u$ into sequence of scene operations $(O_1,\ldots,O_k)$, identifying the resulting set of constraints that should hold.  For each parsed scene operation $O_i$, the system then executes the scene operation by resolving the set of objects on which the operation should be performed and then modifying the scene state accordingly.

\textbf{Command Parsing}
We deterministically map verbs to possible actions as shown in Table ~\ref{tab:sceneOperations}.  Multiple actions are possible for some verbs (e.g., ``place'' and ``put'' can refer to either $Move$ or $Insert$).  To differentiate, we assume new objects are introduced with the indefinite article ``a'' whereas old ones are modified with the definite article ``the''.

\textbf{Object Resolution}
To allow interaction with the scene, we must resolve references to objects within a scene.  Objects are disambiguated by category and view-centric spatial relations.  In addition to matching objects by their categories, we use the WordNet hierarchy to handle hyponym or hypernym referents.  Depending on the current view, spatial relations such as ``left'' or ``right'' can refer to different objects.

\textbf{Camera Positioning}
For the $LookAt$ command, we first identify the set of objects being referred to. Given an utterance $u$ of the form ``look at X'', we get the set of objects that are the most likely referent $Sel = \argmax_{o \in S} P(o|X)$.  We perform a simple viewpoint optimization where we sample camera positions $c$ and find the position that maximizes a view function $f(c)$, giving us $\textit{Cam}_\textit{pos} = \argmax_{c} f(c) $.

Camera positions are sampled from 12 equally spaced points around the up-axis of the selected objects and at a fixed height slightly above the bounding box of the objects.  The camera is targeted at the centroid point of the selected objects.  The view function we use is: $f(c) = \textit{vis}_\textit{Sel}(c) + b(\textit{scr}_\textit{Sel}(c))+ \textit{vis}_\textit{all}(c)$
where $\textit{vis}_\textit{Sel}$ is the number of selected objects visible, $\textit{vis}_\textit{all}$ is the percentage of all objects visible, and $b$ is a function of $\textit{scr}_\textit{Sel}$\footnote{linear ramp from $b(0.2)=0$ to $b(0.4)=1$}, the percent of the screen that is taken up by the selected objects.

\begin{figure}
  \includegraphics[width=\columnwidth]{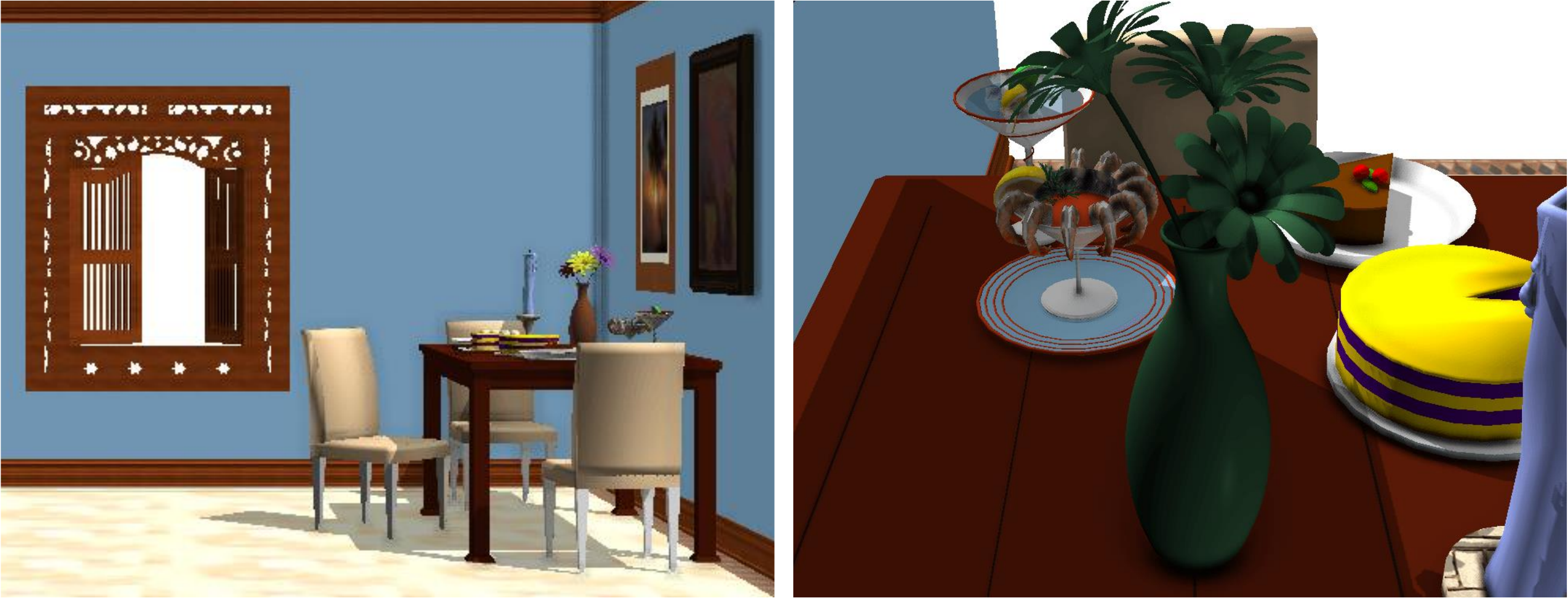}
  \caption{\textbf{Left}: initial scene. \textbf{Right}: after input ``Look at vase'', the camera zooms to the flower vase and the vase is selected (green highlight).}
  \label{fig:lookAtExample}
\end{figure}

This is a simple approach for camera positioning which is challenging to perform manually for novice users of 3D design software and yet critical for navigating and interacting with the environment.

\textbf{Scene Modification}
Based on the operation, we modify the scene by maximizing the probability of a new scene template given the requested action and previous scene template.

\begin{figure}
  \includegraphics[width=\columnwidth]{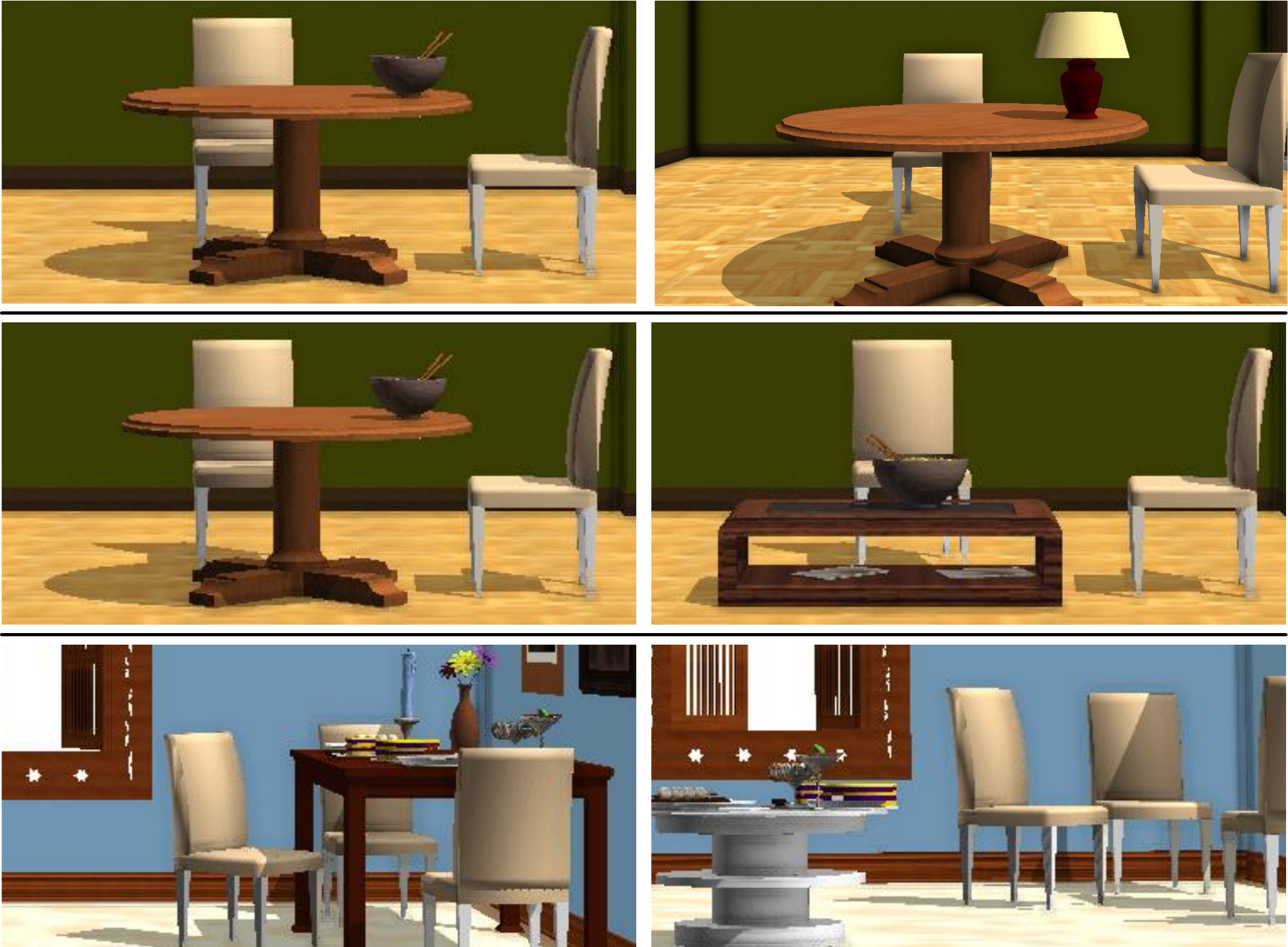}
  \caption{Examples of $Replace$ operation. \textbf{Top}: ``Replace the bowl with a red lamp''. \textbf{Middle}: ``Replace the table with a coffee table''. \textbf{Bottom}: ``Replace the table with a round table''.}
  \label{fig:replace}
\end{figure}

The set of objects in the scene is modified according to the operation:
\vspace{-.5em}
\begin{itemize}[topsep=0pt,noitemsep,parsep=0pt,partopsep=0pt]
\item $Insert$: Select a new object that satisfies the constraints from the model database and place it in the scene.
\item $Replace$: Select a new object that satisfies the constraints from the model database and replace the objects to be replaced with the new object.
\item $Delete$: Remove the old objects from the scene.
\end{itemize}

After the set of objects is determined, relayout is performed on the scene by attempting to satisfy the constraints while minimizing the change in position for all unmodified objects.  For operations such as $Move$ and $Scale$ the set of objects remain the same but their position or size will need to change to satisfy the new constraints.  When doing relayout, we do not vary all parameters, but just the set of influenced parameters for the operation (see  Table~\ref{tab:sceneOperations}).

Depending on the operation, operand objects, and their scene context, the resulting changes in the scene vary in complexity.  The simplest operations only influence the target objects while more complex operations require adjustment of supported objects or even surrounding objects.  Figure~\ref{fig:replace} shows examples of the $Replace$ operation with varying complexity.  Replacing the bowl on the table with a table lamp is relatively easy since the system only needs to ensure the attachment point remains the same.  However, when we replace the table with a coffee table in Figure~\ref{fig:replace} (middle), the bowl needs to be repositioned.  In the bottom row of the figure we see a failure case due to the layout algorithm since the chairs are not repositioned to accommodate the bigger round table. Instead the table is pushed to the side to avoid collisions.  Manually performing a replace operation can be extremely tedious due to such object position dependencies, thus making this a valuable high-level operation.  This illustrates how our approach attempts to bridge the semantic gap between high-level intent and low-level geometric manipulation.  In our implementation, the scene operations are initiated by the user using text commands.  It is also possible to incorporate these operations into a graphical user interface.  For instance, $Replace$ can be implemented in a GUI as a sequence of clicks to select an object, search from a model database, select a desired model to use as a replacement.

\subsection{Scene Refinement}

\begin{figure}
  \includegraphics[width=\linewidth]{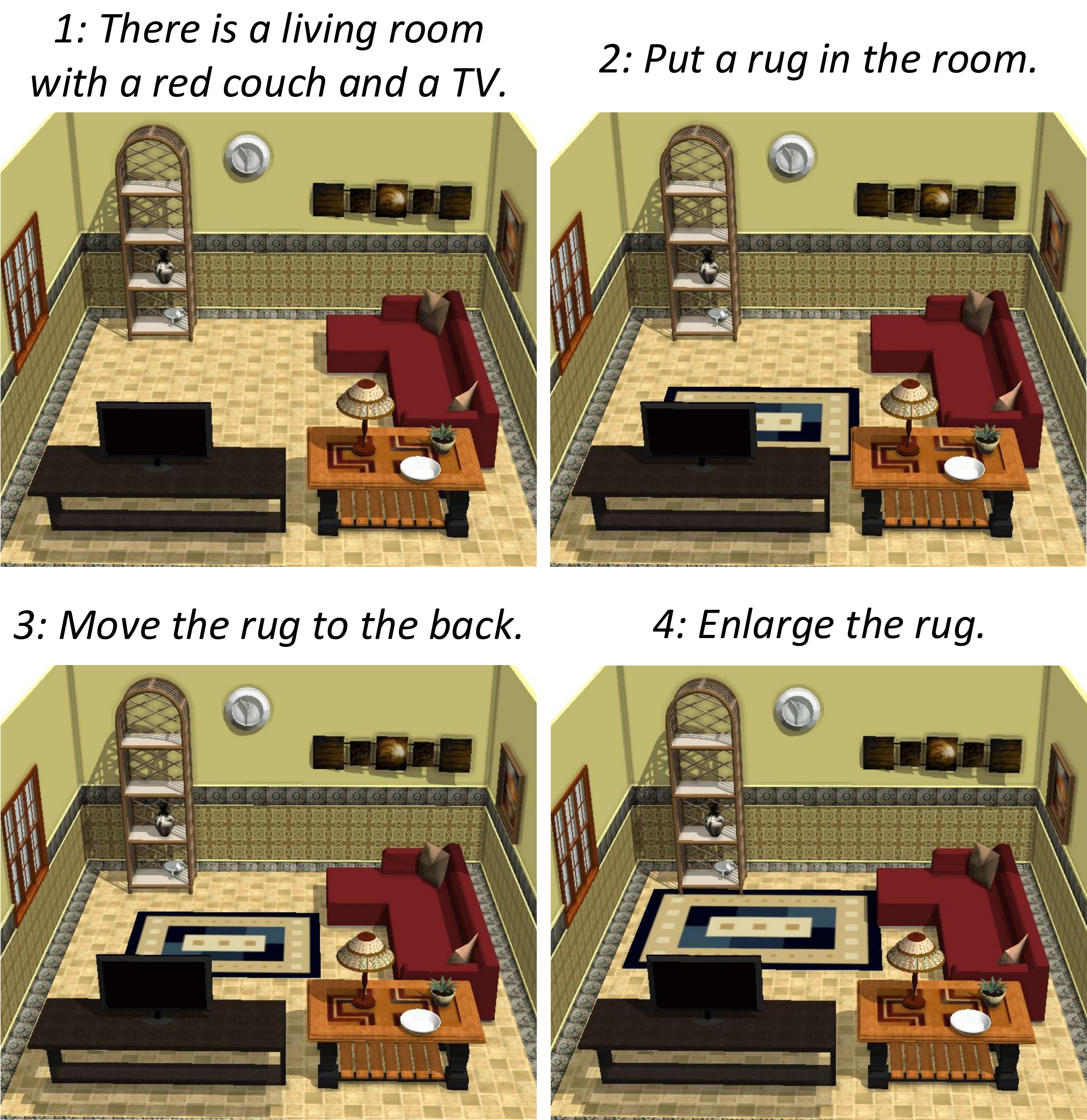}
  \caption{A sequence of textual interactions ro refine a living room scene by adding a rug and manipulating it.}
  \label{fig:rugInteractions}
\end{figure}

With the set of high-level operations that we defined, a user can progressively refine a generated scene.  Figure~\ref{fig:rugInteractions} shows a sequence of operations for adding a rug to the living room we generated earlier.  Combined with traditional mouse and keyboard navigation, these textual interactions allow high-level scene editing operations that are interpreted within the current viewing context (e.g., moving the rug to the back of the room from the viewer's perspective in the third panel).  Without such high-level operations, the necessary interaction would include several steps to find a rug model, place it, orient it, and manually adjust its position.

\begin{figure}
  \centering
  \includegraphics[width=0.7\columnwidth]{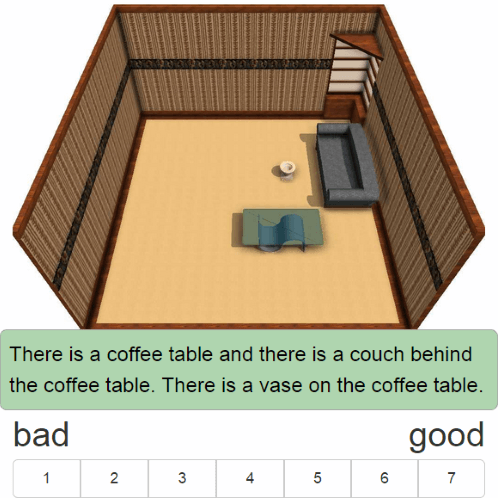}
  \caption{Screenshot from our evaluation experiment. Participants were asked to rate how well the displayed scenes match the description on a 7-point Likert scale.}
  \label{fig:experiment-ui}
\end{figure}

\section{Generated Scene Evaluation}

We evaluate the output of our system by asking people to judge how well generated scenes match given input descriptions.  This is an appropriate initial evaluation since in a practical usage scenario, a scene matching the input description well would provide a good starting point for further refinement.  We compare versions of our system contrasting the benefit of different components. We also establish an upper-bound baseline by asking people to manually design scenes corresponding to the same descriptions, using a simple scene design interface used in prior work~\cite{fisher2012example}.

\begin{figure*}
  \includegraphics[width=\linewidth]{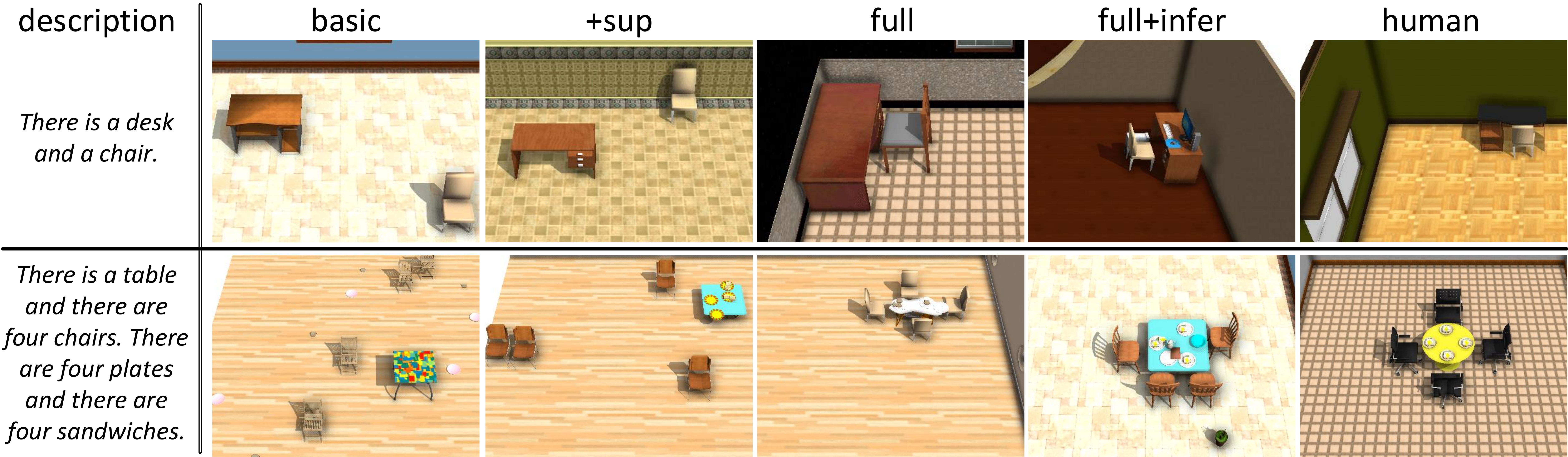}
  \caption{Some example textual descriptions and scenes generated by \SceneSeer in different conditions, as well as scenes manually designed by people.}
  \label{fig:scenesByCondition}
\end{figure*}

\paragraph{Conditions}
We compare seven conditions: \cbasic, \csup, \csupspat, \csupprior, \cfull, \cinfer, and \cmanual.  The \cbasic condition is a simple layout approach which does not use any learned support, spatial relation, or placement priors.  The conditions \csup, \csupspat, and \csupprior, and \cfull (which includes all three priors) test the benefit of adding these learned priors to the system.  Finally, \cinfer performs implicit inference for selecting and laying out additional objects, while \cmanual consists of the manually designed 3D scenes that people created given the descriptions.  For each of these conditions we create a set of 50 scenes, one for each of the input textual descriptions.  In total, we have 350 stimulus scene-description pairs to be rated by people during our experiment (see Figure~\ref{fig:scenesByCondition} for example descriptions and scenes).

\paragraph{Participants}
Participants were recruited online on Amazon Mechanical Turk and were required to be fluent speakers of English.  We recruited a total of 97 participants for evaluating the quality of generated scenes with respect to reference textual descriptions.  For the \cmanual condition, the scenes were created by a different set of 50 participants who were given the textual descriptions and asked to create a corresponding scene (see last column of Figure~\ref{fig:scenesByCondition}).

\paragraph{Procedure}

During the experiment, a randomly sampled set of 30 pairs of generated scene and input textual descriptions were shown to each participant.  The pairs were drawn from all conditions and present in randomized order.  The participants were asked to rate each pair on a 7-point Likert scale to indicate ``how well the scene matches the description'', with a score of 1 indicating a very bad match, and 7 indicating a very good match.  The participants were instructed to consider three points in particular: (1) Are the objects mentioned in the description included in the scene? (2) Are the described relationships between the objects correct in the scene? and (3) Overall, does the scene \emph{fit the description}?  Figure~\ref{fig:experiment-ui} shows a screenshot of the UI that we used to carry out the experiment.

\paragraph{Design}

The experiment was a within-subjects factorial design with the condition \{\cbasic, \csup, \csupspat, \csupprior, \cfull, \cinfer, \cmanual{}\}, description \{1...50\}, and participant \{1...97\} as factors.  The Likert rating for description-to-scene match was the dependent measure.

\subsection{Results}

In total the 97 participants gave 2910 scene ratings for the 350 stimulus scene-description pairs with 8.39 ratings on average per pair (standard deviation was 3.56 ratings).

\begin{figure}
  \centering
  \includegraphics[width=0.8\columnwidth]{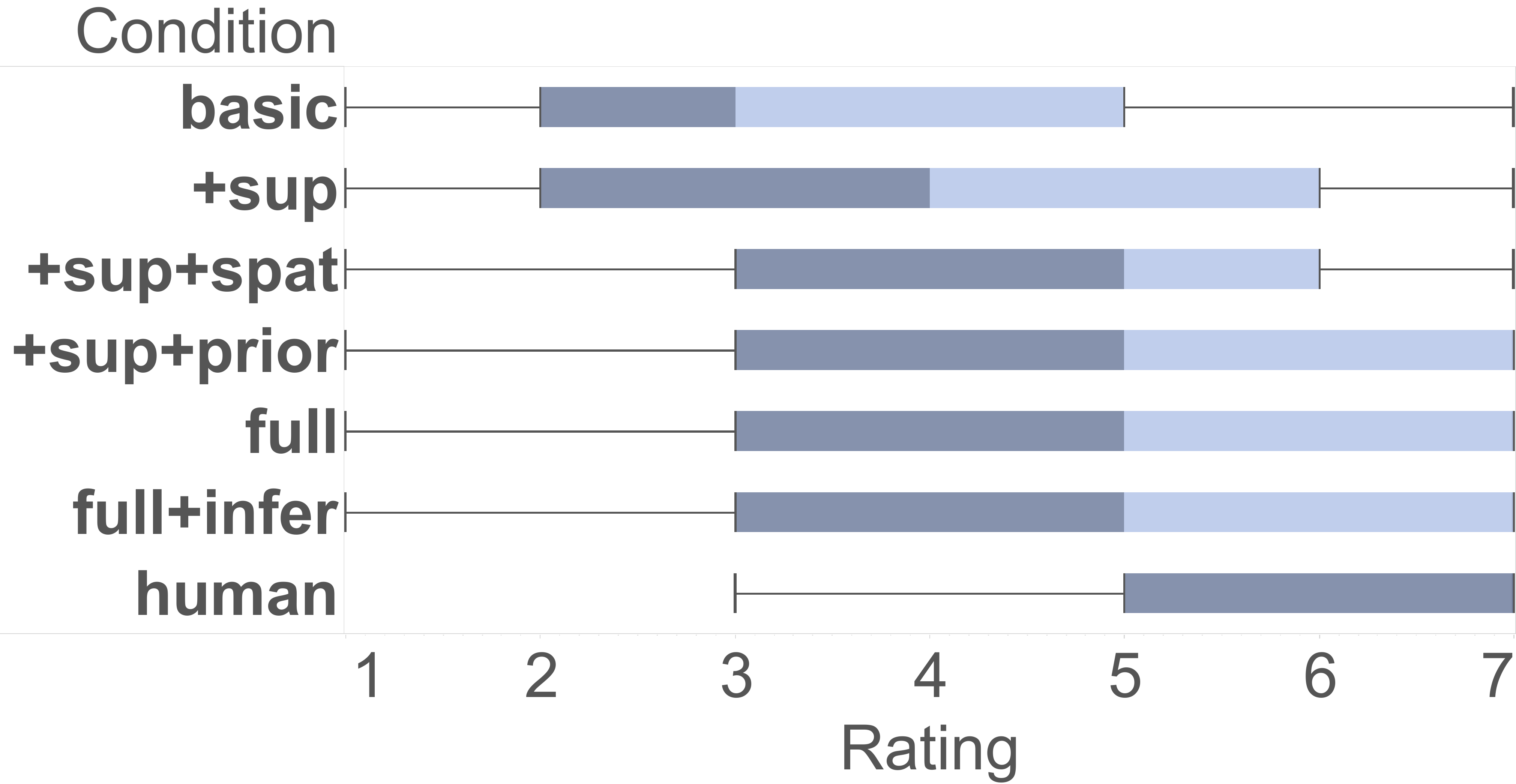}
  \caption{Distributions of scene-description match ratings for experiment conditions. The dark and light regions delineate the lower and upper quartile respectively, and the whiskers extend to 150\% of the interquartile range.}
  \label{fig:ratings-boxplot}
\end{figure}

\paragraph{Generated scene ratings}
The mean ratings for each condition in our experiment are summarized in Table~\ref{tab:humanEvalResults} and the rating distributions are plotted in Figure~\ref{fig:ratings-boxplot}.  We see that the \cbasic condition receives the lowest average rating, while predictably the scenes designed by people receive the highest rating.  Adding learned support, spatial relation parsing, and priors for relative position improve the rating for the scenes generated by our system, and the \cfull combined condition receives the highest average rating.  We note that the scenes generated with additional inferred objects (\cinfer) actually receive a lower rating.  We hypothesize that two factors may contribute to the lower rating for \cinfer. Firstly, adding extra objects makes the scene layout more complicated and prone to exhibiting object selection or spatial relation errors.  Secondly, inferred objects are not explicitly mentioned in the description so participants may not have viewed them as significantly improving the quality of the scene (though they were instructed to consider additional objects positively if they add naturally to the appearance of the scene).

\begin{table}
  \centering
  \begin{tabular}{lc}
    \toprule
    condition  &  mean rating  \\ \midrule
    \cbasic    &     3.61      \\
    \csup      &     4.22      \\
    \csupspat  &     4.72      \\
    \csupprior &     4.90      \\
    \cfull     & \textbf{5.03} \\
    \cinfer    &     4.65      \\ \midrule
    \cmanual   &     5.89      \\ \bottomrule
  \end{tabular}
  \caption{Average scene-description match ratings for each of the conditions in our experiment.}
  \label{tab:humanEvalResults}
\end{table}

\paragraph{Statistical analysis}
The statistical significance of the effect of the condition factor on the ratings was determined using a mixed effects model, with the condition as a fixed effect and the participant and description as random effects, since the latter two are drawn randomly from a large pool of potential participants and descriptions\footnote{We used the \stt{lme4} R package and optimized for maximum log-likelihood~\cite{baayen2008mixed}.}. Most pairwise differences between conditions for mean scene rating were significant under Wilcoxon rank-sum tests with the Bonferroni-Holm correction ($p\textless0.05$).  The exceptions are the comparisons between \csupspat, \csupprior and \cfull which were not significant.

\paragraph{Summary}
The experimental results show that \SceneSeer can generate plausible 3D scenes given input descriptions.  The different components of the system that leverage learned support relations, spatial relationship parsing, and relative position priors all contribute towards improving the quality of the generated scenes.  Implicitly inferred additional objects do not improve the scene-to-description match ratings given by people.

\section{Discussion}

\subsection{Limitations}

The unoptimized implementation of our system takes several seconds to parse, infer and generate a scene, depending mostly on the number of objects and relations expressed in the input.  Unfortunately, this makes interactive use slow in practical settings.  However, the system can be easily optimized to within the range of real-time editing.  A more important limitation, is that we currently handle a limited domain of scenes.  Parsing of scene descriptions with existing natural language processing methods is error-prone even for simple, everyday language, and the publicly available 3D model datasets are biased towards furniture and electronics, limiting the types of potential output scenes (e.g., ``the flower garden outside the window'' is a challenging case).  Furthermore, generation of scenes with more organic, complex object geometry is challenging for our simplistic layout approach, and we have ignored the importance of stylistic consistency in object selection (e.g., different styles of chairs selected for same table).

The current set of operations that we presented is constrained to fairly simple manipulations of the scene.  An improved system would be able to process more complex interactions that require a deeper semantic understanding of the world.  For example: ``move the TV to face the couch'' (requires knowledge of front sides and facing orientations), and ``put a bowl of fruit on the second shelf of the bookcase'' (requires knowledge of object parts and relative ordering from bottom).

We have not explored the design choices in combining a high-level natural language interface such as \SceneSeer with direct manipulation is an interesting avenue for future research.  Different operations might be more efficient to perform in one modality or the other, depending also on the particular scene context and viewpoint.  Measuring the relative benefits of different combination strategies in specific tasks is an interesting avenue for future research.
\vspace{-.1em}

\subsection{Future Work}

A promising direction for future work is to integrate systems such as \SceneSeer with speech to text systems allowing for voice-driven and more natural dialogue interactions.  This could be particularly relevant in the context of increasingly popular virtual and augmented reality systems where text input is more cumbersome.

Leveraging the interactive nature of the system to improve spatial knowledge is an exciting avenue for future work.  For instance, by observing where the user decides to manually place objects, we can improve our placement priors.  A natural way for the user to provide feedback and corrections to system errors is through a dialogue-based interaction extending the imperative textual commands.

Another exciting direction for future work is to crowdsource the accumulation of spatial knowledge by scaling the system on web-based platforms.  This will also provide an opportunity for broader user studies that can give insight into context-specific preferences for text versus direct manipulation interactions, and provide useful data for informing the design of future text-to-scene systems.

\section{Conclusion}

We presented \SceneSeer, a text to 3D scene generation system with semantically-aware textual interactions for iterative 3D scene design.  We evaluated the scenes generated by our system through a human judgment experiment and confirmed that we can generate scenes that match the input textual descriptions and approach the quality of manually designed scenes.  We demonstrated that automatically generated 3D scenes can be used as starting points for iterative refinement through a set of high-level textual commands.  Natural language commands are a promising mode of interaction for 3D scene design tasks since they abstract much of the tedium of low level manipulation.  In this way, we can bridge the semantic gap between high-level user intent and low-level geometric operations for 3D scene design.

We believe that the domain of text to 3D scene design presents many challenging research problems at the intersection of computer graphics, HCI, and natural language processing.  Our system presents a small step in the direction of enabling natural language interfaces for scene design.  We believe that in the future, dialogue-based systems will make the scene design process as effortless as telling a story.

\bibliographystyle{acm-sigchi}
\bibliography{scenegen}

\appendix

\balance

\subsection{Scene Layout Score}
\label{app:layoutScore}

The scene layout score is given by $\mathcal{L} = \lambda_\textit{obj} \mathcal{L}_\textit{obj} + \lambda_\textit{rel} \mathcal{L}_\textit{rel}$, a weighted sum of object arrangement $\mathcal{L}_\textit{obj}$ score and constraint satisfaction $\mathcal{L}_\textit{rel}$ score following from Chang et al.~\cite{chang2014spatial}'s definition:
\begin{eqnarray*}
  \mathcal{L}_\textit{obj} &=& 
  \sum_{o_i} P_\textit{surf}(S_n|C_{o_i}) \sum_{o_j \in F(o_i) }P_\textit{relpos}(\cdot)\\
  \mathcal{L}_\textit{rel}&=&
  \sum_{c_i} P_\textit{rel}(c_i)
\end{eqnarray*}
where $F(o_i)$ are the sibling objects and parent object of $o_i$.  We use $\lambda_\textit{obj} = 0.25$ and $\lambda_\textit{rel} = 0.75$ for the results we present.

\subsection{Spatial Knowledge Priors}

We use similar definitions for spatial knowledge priors as those presented by Chang et al.~\cite{chang2014spatial} with updated support and attachment surface priors.  Spatial knowledge priors are estimated using observations in scenes. To handle data sparsity we utilize our category taxonomy.  If there are fewer than $k=5$ support observations we back off to a parent category in the taxonomy for more informative priors.

\paragraph{Object Occurrence Priors}
Occurrence priors are given by simple Bayesian statistics of objects occurring in scenes: $P_\textit{occ}(C_{o}|C_{s}) = \frac{\text{count}(C_{o} \text{~in~} C_{s})}{\text{count}(C_{s})}$

\paragraph{Support Hierarchy Priors}
We estimate the probability of a parent category $C_p$ supporting a given child category $C_c$ as a simple conditional probability based on normalized observation counts: $P_\textit{support}(C_{p}|C_{c}) = \frac{\text{count}(C_{c} \text{~on~} C_{p})}{\text{count}(C_{c})}$

\paragraph{Support and Attachment Surface Priors}
Similarly, the parent support surface priors are given by: $P_{\textit{surf}_\textit{sup}}(s | C_c) = \frac{\text{count}(C_c \text{~on surface with~} s)}{\text{count}(C_c)}$
The parent supporting surface is featurized using the surface normal (up, down, horizontally) and whether the surface is interior (facing in) or exterior (facing out).  For instance, a room has a floor which is an upwards interior supporting surface: 

The child attachment surface priors are given by: $P_{\textit{surf}_\textit{att}}(s | C_c) = \frac{\text{count}(C_c \text{~attached at surface~} s)}{\text{count}(C_c)}$
Object attachment surfaces are featurized using the bounding box side: top, bottom, front, back, left, or right.  For instance, posters are attached on their back side to walls, rugs are attached on their bottom side to floors.

If there are no observations available we use the model geometry to determine the support and attachment surface. For support surfaces we pick only upward facing surfaces, while for attachment we assume 3D (blocky) objects are attached on the bottom, flat objects are attached on their back or bottom, and thin objects are attached on their side.

\paragraph{Relative Position Priors}
We model the relative positions of objects based on their object categories and current scene type: i.e., the relative position of an object of category $C_\textit{obj}$ is with respect to another object of category $C_\textit{ref}$ and for a scene type $C_{s}$.  We condition on the relationship $R$ between the two objects, whether they are siblings ($R=\textit{Sibling}$) or child-parent ($R=\textit{ChildParent}$), and define the relative position prior as: $P_\textit{relpos}( x,y,\theta | C_\textit{obj}, C_\textit{ref}, C_{s}, R )$ which we estimate using kernel density estimation.

\end{document}